# Student Difficulties in Translating between Mathematical and Graphical Representations in Introductory Physics

Shih-Yin Lin, Alexandru Maries and Chandralekha Singh

*Department of Physics and Astronomy, University of Pittsburgh, Pittsburgh, PA 15260*

**Abstract.** We investigate introductory physics students' difficulties in translating between mathematical and graphical representations and the effect of scaffolding on students' performance. We gave a typical problem that can be solved using Gauss's law involving a spherically symmetric charge distribution (a conducting sphere concentric with a conducting spherical shell) to 95 calculus-based introductory physics students. We asked students to write a mathematical expression for the electric field in various regions and asked them to graph the electric field. We knew from previous experience that students have great difficulty in graphing the electric field. Therefore, we implemented two scaffolding interventions to help them. Students who received the scaffolding support were either (1) asked to plot the electric field in each region first (before having to plot it as a function of distance from the center of the sphere) or (2) asked to plot the electric field in each region *after* explicitly evaluating the electric field at the beginning, mid and end points of each region. The comparison group was only asked to plot the electric field at the end of the problem. We found that students benefited the most from intervention (1) and that intervention (2), although intended to aid students, had an adverse effect. Also, recorded interviews were conducted with a few students in order to understand how students were impacted by the aforementioned interventions.

## INTRODUCTION

A solid grasp of different representations of knowledge in a given domain, e.g., verbal, mathematical and graphical, including the facility with which one can transform knowledge from one representation to another is a hallmark of expertise [1-3]. Physics experts automatically transform problems from the initial representation into a representation more suitable for further analysis in attempting to solve problems. However, introductory physics students not only need explicit help understanding that choosing an appropriate representation is an important step in organizing and simplifying the given information, but they also need help in learning to transform knowledge from one representation to another [4]. Here, we explore the facility of students in a calculus-based introductory physics course in transforming from mathematical to graphical representation a problem solution involving the electric field for spherical charge symmetry.

## METHODOLOGY

A class of 95 calculus-based introductory physics students was enrolled in three different recitations. The three recitations formed the comparison group and two intervention groups for this investigation. In addition, six students in a different but equivalent calculus-based physics class were interviewed using a think-aloud protocol to understand their thought processes better while they solved the problem. Below, we first describe the interventions used in two of the recitations. All recitations were taught in a traditional way in which the TA worked out problems similar to the homework problems and then gave a 15-20 minute quiz at the end of class. Students in all recitations attended the same lectures, were assigned the same homework, and had the same exams and quizzes. Here we analyze students' difficulties in transforming the solution to the following problem from the mathematical to the graphical representation after they worked on it in a quiz.

"A solid conductor of radius *a* is inside a solid conducting spherical shell of inner radius *b* and outer radius *c*. The net charge on the solid conductor is +$Q$ and the net charge on the concentric spherical shell is –$Q$ (see figure).

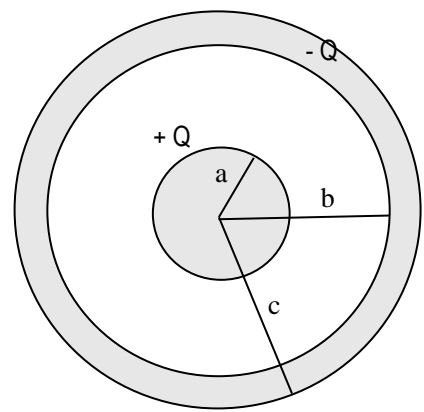

(a) Write an expression for the electric field in each region.
  (i)   $r < a$
  (ii)  $a < r < b$
  (iii) $b < r < c$
  (iv)  $r > c$
(b) On the figure below, plot $E(r)$ (which is the electric field at a distance *r* from the center of the sphere) in all

regions for the problem in (a)." (they were given coordinate axes on which to plot the electric field)

Our previous preliminary research in a different introductory calculus-based physics class suggested that students have difficulty graphing the electric field after writing an expression for the electric field in each region. In particular, a majority of students (~70%-80%) drew graphs that were not consistent with their mathematical expressions in all regions. Motivated by our preliminary findings, we implemented two interventions in two of the recitations (which will be referred to as Group 1 and Group 2) by giving students some scaffolding support in order to assess if it helps them make a connection between the two representations. The problem above without additional scaffolding was given to the third recitation which will be referred to as the comparison group or "Group 3". Theoretical task analysis of the process of transforming from mathematical to graphical representation was used to design the two interventions. The first intervention group (Group 1) was asked to plot the electric field in each region before graphing it in part (b) shown above. Their instructions were the following:

"(a) Write an expression for the electric field in each region and plot the electric field in that region on the coordinate axes shown (in the shaded region, please do not draw)." They were then given the coordinate axes shown below where the irrelevant parts were shaded out so that students would readily recognize the region where they are required to plot the electric field.

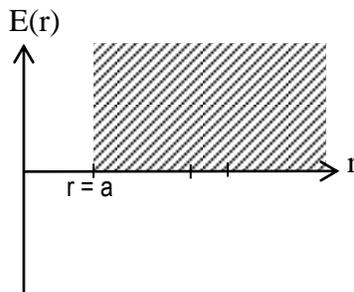

**Figure 1:** Coordinate axes given in region $r<a$ to students in Group 1

Students in the second intervention group (Group 2) were asked to plot the electric field in each region after calculating the field in that region and, similar to Group 1, they were given coordinate axes with the irrelevant regions shaded out. However, in addition, they were asked to find the electric field at the beginning, mid and end points of each region before graphing it in that region. For example, for region $r<a$, they were also asked to fill in the following blanks after writing an expression for the electric field, but before graphing it:

When $r=0$, $E(r=0)$ = __________

When $r = \dfrac{a}{2}$, $E(r = \dfrac{a}{2})$ = __________

When $r \to a$, $E(r \to a)$ = __________

Both interventions (which were given to Groups 1 and 2) were designed to help students perform better on graphing the electric field. We hypothesized that asking students to graph the field in each region first after writing an expression for the field in that region but before constructing the graph for the field everywhere may help them in making a connection between the graphical and mathematical representations better. In particular, we hoped that students would realize that in this problem the electric field takes the form of a piece-wise defined function (with discontinuity in the electric field when one crosses a surface charge distribution) and in order to graph it they need to match the forms of this function in the corresponding regions. The extra instructions for intervention Group 2 to find the electric field at the beginning, mid, and end point of each region before graphing in that region were intended to, on the one hand, give another hint that the electric field has different forms in different regions, and on the other hand, help students see that the electric field has discontinuities at interfaces carrying surface charges and thus help them perform better on graphing it.

The researchers jointly determined the grading rubric. After discussions among the researchers, the way the problem was scored is shown in Table 1.

**TABLE 1:** Points assigned to each part of the problem.

| Region | Part a | Part b |
|---|---|---|
|  | Expression for the electric field | Plot of $E(r)$ in region |
| (i) $r < a$ | 10 points | 5 points |
| (ii) $a < r < b$ | 30 points | 15 points |
| (iii) $b < r < c$ | 10 points | 5 points |
| (iv) $r > c$ | 10 points | 5 points |

Table 1 shows that the region $a<r<b$ was worth three times more than each other region. The reason for this choice is because this region was the only one with a non-zero electric field. In finding the expression for the electric field in parts (a)(i) through (a)(iv), students were given 80% for the correct expression and 20% for the correct reasoning that led to that expression. Table 1 also shows that graphing the electric field in part (b) was worth 30 points which is half of the points assigned to finding the expressions for the electric field in part (a). Part (b) was broken up into individual regions and in each region we investigated if the students' chosen graph was consistent with the expression found for the electric field in that region. We were only interested in the form of the graph matching the expression; the students did not need to label endpoints, or even have correct endpoints. For example, if a student found $E(r) = kr/3$ in region $b<r<c$, and drew an increasing linear graph that starts from the $r$ axis; this student would be considered to be consistent (and get the 5 points assigned to this part) because he/she selected the correct type of graph (linear) consistent with the expression in that region, although the left endpoint is clearly incorrect (based on the expression, $E(r=b) \neq 0$). We note that students in Groups 1 and 2 did not obtain any extra points for graphing the electric field in each region first, or for finding the electric field at the beginning, mid and end points of each region.

## QUANTITATIVE RESULTS

Before presenting the results it is worthwhile to mention that we used students' scores on the final exam (as graded by their instructor) to ensure that the groups performed about the same. There was no statistical difference in the performance of students in different groups on the final exam.

Table 2 shows the averages of each group in the parts that required finding an expression for the electric field. *T*-tests on data in Table 2 reveal that students in Group 1 outperformed students in Group 3 in part (a) (ii) (p = 0.022, Cohen's d = 0.628) and in part (a) (iii) (p = 0.019, Cohen's d = 0.654). Also, students in Group 1 outperformed students in Group 2 in part (a) (i) (p = 0.006, Cohen's d = 0.816) and in part (a) (iii) (p = 0.005, Cohen's d = 0.843). We note that students in Group 1 have better averages than students in Groups 2 and 3 by at least 20% in *every part* (see Table 2) but the differences are only statistically significant in the cases noted above.

**TABLE 2:** Averages of each group in the parts that required finding an expression for the electric field out of 10 points (score in region $a<r<b$ was renormalized to 10 maximum points).

| Group | $r<a$ | $a<r<b$ | $b<r<c$ | $r>c$ |
|-------|-------|---------|---------|-------|
| 1 | 5.7 | 5.7 | 5.4 | 5.9 |
| 2 | 2.8 | 4.7 | 2.0 | 3.4 |
| 3 | 4.3 | 3.4 | 2.8 | 3.4 |

We also investigated if students in some groups were more consistent than others. Table 3 shows the percentage of students from each group who were *always* consistent; that is, they were consistent in all the graphs they drew. Chi-squared tests on the data shown in Table 3 show that students in Group 1 are more consistent than students in the other two groups (p = 0.005 for comparison to Group 2 and p = 0.006 for comparison to Group 3).

**TABLE 3:** Percentages (and numbers) of students in each group who were consistent in all parts

|  | Consistent in all parts | |
|--|---|---|
|  | Yes | No |
| Group 1 | 62% (16) | 38% (10) |
| Group 2 | 23% (6) | 77% (20) |
| Group 3 | 26% (8) | 74% (23) |

It is interesting to note that the only difference between Groups 1 and 2 is that students in Group 2 were also asked to find the electric field at the beginning, mid and end points of each region before graphing it. This extra instruction which was intended to help students graph better had adverse effects, making them *less* consistent. This is not at all what was expected from a theoretical task analysis and we could only come up with a possible explanation after conducting individual interviews with some students who solved the version of the quiz given to Group 2.

## QUALITATIVE RESULTS

Interviews provided a possible explanation for students' poor performance in Group 2. One cognitive framework that can explain the poor performance of students in Group 2 pertains to working memory (short term memory or STM). In this framework, problems are solved by processing relevant information in the STM and STM is finite (5-9 "slots") for a person regardless of their intellectual capabilities. In order to solve a problem one has to process relevant information in STM in order to move forward with a solution. Novices are more likely to have cognitive overload while solving problems if there are too many smaller chunks of information to keep track of in STM during problem solving. Moreover, since novices do not have a robust knowledge structure, they are more likely to focus on unimportant features of the problem and get distracted.

Students in Group 2 had the extra instructions to find the electric field at the beginning, mid and endpoints of each interval. A cognitive task analysis from an expert point of view suggests that these are good things to calculate when graphing a function because they give information about the function explicitly which is helpful for graphing it. However, the interviews suggested that Group 2 students did not discern the relevance of these instructions to graphing the function in the next part and they had cognitive overload due to the additional instructions. In particular, interviews suggested that these students were more likely to lose track of important, relevant information and even omitted reading instructions carefully. *Every single student interviewed* did not read the instructions carefully at some point. Some forgot to graph the electric field in a particular region, some went straight to the limits ($E(r\rightarrow a)$, $E(r\rightarrow b)$ etc.) even before finding an expression for the electric field in that region. An interesting example of losing track of important information comes from an interview with John. In finding the limits of the function in regions $r<a$ and $a<r<b$ John did not plug into the corresponding values for $r$. For example, he wrote $E(r\rightarrow a) = kQ/r^2$ without plugging $r = a$ in the expression. But then when he got to the first limit in region $b<r<c$ ($E(r\rightarrow b)$), after writing down an initial expression in which he did not plug in $r=b$, he suddenly realized, without the interviewer (Int) saying anything, that he should plug in $r = b$:

John: "Oh, should I plug in […] 'cause it's r approaching b?"
Int: "I can't tell you that. […] What do you think?"
John: "I'll just write it to be safe."

He then went back and changed all the previous limits where he had not plugged in the corresponding values for $r$. Thus, the piece of information "when you find a limit of a function, you have to plug in the value for the variable in that function" was present in his long term memory but he did not retrieve it until a particular point. He appeared to be focusing on and processing other information in the problem that was not helpful for figuring out the limits correctly. As noted earlier, *every single student interviewed* overlooked something in a somewhat similar manner while solving the different parts of the problem and the intended scaffolding involving explicit evaluation of the function at three points in each region did not help them in transforming the expression for the field in a region to the graphical representation correctly.

## DISCUSSION and SUMMARY

So far we discussed a possible explanation for the poor performance of Group 2 (as compared to Group 1), but we have not discussed a possible reason that can explain why students in Group 1 performed better than the comparison group (Group 3) both in terms of consistency and score. We note that the difference between the two groups is that Group 1 was also asked to plot the electric field in each region immediately after writing an expression for it, but before plotting it again in all the regions at the end. For each region, they were also given coordinate axes with the irrelevant regions shaded out to help them isolate the relevant information in the problem. Part of the reason why students in Group 1 were more likely to be consistent may be that some of them benefited from having to plot the electric field immediately after finding an expression by realizing (as an expert would) that the electric field has different definitions in each region which need to be plotted independently of one another. The intervention given to Group 1 was meant to aid them in graphing and it appears to be doing just that. However, not only are students in Group 1 more consistent in graphing, but they are also obtaining better scores in finding the electric field in part (a). This can be surprising because the scores in part (a) are based on how successful students are in obtaining the correct expression for the electric field; graphing it was not included in the score. One theoretical model to make sense of this once again invokes STM and cognitive load. Students' knowledge chunks are smaller than an expert's and sometimes students may focus on pieces of information that are not necessarily helpful for solving the problem which in turn can cause cognitive overload. It is possible that giving students the picture with coordinate axes with only the region at hand unshaded helped them focus on that particular region and some students benefited from this by ignoring irrelevant information. For example, when solving for the electric field in regions $r<a$, $a<r<b$ and $b<r<c$ one has to ignore any contributions coming from the $-Q$ charge on the outer shell.

We found that asking students to plot the electric field in each region immediately after finding an expression for it in that region impacted students positively, making them more likely to find the correct expressions and more likely to be consistent in plotting the expressions they found. We hypothesize that giving students coordinate axes with the irrelevant regions shaded out may have focused their attention on the relevant information in the problem that needed to be taken into account while finding the expressions for the electric field and graphing them. This would make students more effective at solving the problem and could partly account for their improved performance.

We also found that the added instructions to evaluate the electric field at the beginning, mid, and endpoints of each interval, although intended to help students be more consistent in graphing the electric field, had an adverse effect on their performance (both in terms of score and consistency). Conducted think-aloud interviews with students suggested that they did not discern the relevance of these instructions and ended up having cognitive overload due to the added information that needed to be processed in STM while engaged in solving this problem.

## REFERENCES

1. J. Larkin and H. Simon, Cognitive Science, **11**, 65-99 (1987).
2. J. Zhang, Cognitive Science, **21**, 179-217 (1997).
3. A. Newell and H. Simon, *Human Problem Solving*, Engelwood Cliffs NJ: Prentice Hall, 1972
4. L. McDermott, M. Rosenquist and E. van Zee, Am. J. Phys., **55**, 503-513 (1987).